\documentclass[prl, groupedaddress, showpacs, reprint, floatfix]{revtex4-1}
	\usepackage{graphics}
	\usepackage{graphicx}
	\usepackage{textcomp}
	\usepackage{amssymb}
	\usepackage{amsmath}
	\usepackage{color}
	\usepackage[hypertex]{hyperref}
\begin{document}
\title{Single Flux Transistor: the controllable interplay of Coherent Quantum Phase Slip and Flux quantization}

\author{S.~Kafanov$^{1}\footnote{Electronic address:
sergey.kafanov@gmail.com}$, N.\,M.~Chtchelkatchev$^{2, 3}$}

	\affiliation{$^1$ The Institute of Physical and Chemical Research (RIKEN), 34 Miyukigaoka, Tsukuba, Ibaraki, 305-8501, Japan}
	
	\affiliation{$^2$Institute for High Pressure Physics, Russian Academy of Science, Troitsk 142190, Russia}
	
	\affiliation{$^3$ Department of Theoretical Physics, Moscow Institute of Physics and Technology, 141700 Moscow, Russia}
	


\begin{abstract}
	The Single Cooper Pair Josephson Transistor is a device that exhibits at the same time charge quantization and phase coherence. Coherent quantum phase slip phenomenon is "dual" the Josephson phase coherence while the charge quantization is dual to the flux quantization. We present the experimental demonstration and the theoretical description of a new superconducting device -- Single Flux Transistor, which is dual to the Single Cooper Pair Transistor. Our transport measurements show the periodic modulation of the \textit{critical voltage} by the external magnetic field. The obtained current-voltage characteristics show the hysteretic behavior, which we attribute to the intrinsic self-heating of charge carriers.
\end{abstract}
\maketitle

Superconductivity, when the charge carriers move coherently thus allowing electrical current without any voltage drop, is one of the central topics of research in condensed matter physics \cite{Phys.Rev.108.1175.Bardeen,PhysLett.1.251.Josephson}. The analog of this effect is a coherent motion of magnetic fluxes, so called the coherent quantum phase slips (CQPS)\cite{NaturePhysics.2.169.Mooij}. This motion generates the electrical voltage without any driving current. The phase and the particle number operators do not locally commute that makes the effects of superconducting coherence and charge quantization competitive with each other. However, these effects may coexist when moved apart at the nanoscale in the single Cooper pair transistor, where the supercurrent depends on the charge state of the transistor, and can also be controlled by an external gate voltage. Hereby we present the experimental and theoretical analysis of the dual physical system: coherent quantum phase slip single flux transistor, in which coherent motion of quantized magnetic fluxes is controlled by an external magnetic field.

The phase slip effect is well known in superconducting physics. It appears in  low-dimensional superconductors, which demonstrate a finite resistance even at temperatures substantially below the superconducting transition. At temperatures close to the superconducting transition this resistance is caused by thermally activated phase slips \cite{PhysRevLett.20.461.Webb}. Incoherent quantum flux tunneling is an origin of the finite resistance at lower temperatures for quasi-1D structures, i.e. nanowires \cite{PhysRevLett.61.2137.Giordano, PhysRevLett.74.5128.Duan, PhysRevLett.78.1552.Zaikin, Nature.404.971.Bezryadin}. A scheme for demonstrating CQPS was theoretically proposed \cite{NewJournalPhysics.7.219.Mooij}, and experimentally realized \cite{Nature.484.355.Astafiev} recently in a new type of qubit. However, in spite of a number of theoretical proposals \cite{PhysRevB.83.174511.Hriscu} the unambiguous observation of this phenomenon in transport measurements has not been reported so far.

\begin{figure}[b]
\centering
\includegraphics[width = \linewidth]{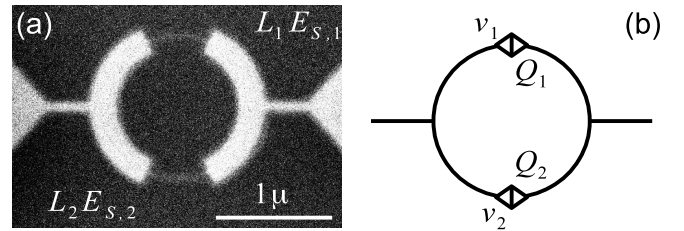}
\caption{\label{fig:Scheme} (a) Scanning electron micrograph of a typical device used in our experiments. It is a superconducting $\mathrm{In O_x}$ ring with a $\mathrm{1\mu m}$ inner diameter. Two parts of the ring were substituted by a $\mathrm{40\,nm}$ wide and $\mathrm{0.5\,\mu m}$ long quasi-1D wire made of the same material. The nanowires are characterized by the phase-slip coupling energies $E_{S_{1,\,2}}$, and kinetic inductances $L_{1,\,2}$. (b) SFT circuit. The device includes two QPS elements denoted by diamond symbols. Each CQPS junction is characterized by the charge difference across the junction $Q_i$ and the number of fluxes $\nu_i$ that tunnel through these junctions.}
\end{figure}
The SEM image of our device is shown in Fig.\,\ref{fig:Scheme}(a). Our device is a superconducting ring, which is incorporated in a measurement scheme via two superconducting probes at diametrically opposite points on the ring. Two parts of the ring are replaced with superconducting nanowires, which allows fluxes to tunnel in and out of the ring. An external magnetic field can be applied perpendicular to the plane of the ring. This device demonstrates the periodic modulations of the \textit{phase-slip blockade} as a function of the external magnetic field (that plays the role of the "gate") with a period equal to the magnetic flux quantum $\Phi_0=h/(2e)$. This behavior is dual to the gate-controlled modulation of the critical current in a Cooper-pair transistor (CPT) \cite{Averin.Likharev, PhysRevLett.70.2940.Matveev, PhysRevLett.72.2458.Joyez}, and therefore call our device a single flux transistor (SFT). We should point out that the ring-patterned geometry experiments have already been reported, however, the flux dynamics in these experiments was governed by incoherent flux tunneling \cite{PhysicaB.227.235.Kanda, Sci.Rep.2.293.Arutyunov}.

The scheme of our device is shown in Fig.\ref{fig:Scheme}. The device was modeled using two CQPS junctions connected in parallel. Each branch $i$ of the ring has an internal inductance $L_i$, and a CQPS junction with a coupling energy $E_{S_{i}}$. The external magnetic field is described by the magnetic flux $\Phi$ through the device. The CQPS junctions can be naturally characterized by a pair of canonically conjugate variables, i.e. the charge difference across the junction $Q_i$ normalized by $2e/2\pi$ and the number of flux quanta $\nu_i$, that tunnel across this junction. Each flux tunneling event is associated with a phase-slip process, when the superconducting phase difference along the nanowire changes by $2\pi$. It will be convenient to introduce a new set of variables, namely, the sum and differences of the variables of the individual junctions. The $\rho=Q_1+Q_2$ value is associated with an electric current through probe contacts $I=(e/\pi)\dot{\rho}$. The conjugate variable to $\rho$ is the total number of fluxes that have tunneled through both CQPS junctions $\nu=(\nu_2+\nu_1)/2$. The second pair of conjugate variables is the number of excess fluxes in the loop $\hat{n}=\nu_1-\nu_2$ and $\hat{q}=(Q_2-Q_1)/2$, which describes the quantum charge fluctuations in the ring. The Hamiltonian of the SFT is given by:
\begin{equation}
	\label{eqn:Hamiltonian1}
	\hat{H} = E_L(\hat{n}-\Phi/\Phi_0)^2 - E_{S}(\rho)\cos\left(\hat{q}+\chi(\rho)\right)\,,
\end{equation}
where $E_L=\Phi_0^2/(2L_\Sigma)$ is the magnetic energy of the ring, corresponding to the total ring inductance $L_\Sigma=L_1+L_2$, where $L_{1,2}$ are the kinetic inductances of each branch of the ring. The second term in the Hamiltonian describes the effective phase-slip coupling energy and contains the phase-slip energies of both junctions:
\begin{align}
	& E_S(\rho)=\sqrt{E_{S_1}^2+E_{S_2}^2+2E_{S_1}E_{S_2}\cos(\rho)}\,,\\
	\nonumber & \chi(\rho) = \arctan\left(\frac{E_{S_1}-E_{S_2}}{E_{S_1}+E_{S_2}}\tan \frac{\rho}{2}\right)
\end{align}
If bias $V_b$ is applied to the SFT, an extra term $-(e/\pi)\rho V$ is also added to the Hamiltonian. In a symmetric case, when $E_{S_1}=E_{S_2}$ the Hamiltonian (\ref{eqn:Hamiltonian1}) is simplified to
\begin{equation}
	\label{eqn:Hamiltonian2}
	\hat{H}=E_L(\hat{n}-\Phi/\Phi_0)^2-E_S(\rho)\cos\hat{q}\,.
\end{equation}
This Hamiltonian is similar to the Hamiltonian of a single CQPS junction \cite{NaturePhysics.2.169.Mooij}. However, the first term of the SFT Hamiltonian has the experimentally controlled parameter $\Phi/\Phi_0$, and the effective coupling energy $E_S=2E_{S_{1}}\cos(\rho/2)$ depends on the charge difference across the SFT. The Hamiltonian (\ref{eqn:Hamiltonian2}) can be diagonalized in a subspace of flux states $|n\rangle$:
\begin{align}
	H &= \sum_{n \in \mathbb{Z}} \left\{\right.E_L(n-\Phi/\Phi_0)^2 |n\rangle \langle n|\\
	\nonumber &- \frac{1}{2}E_S(\rho)\left(|n\rangle\langle n+1|+|n+1\rangle\langle n|\right)\left.\right\}.
\end{align}
The eigen energy levels $E_m(\rho,\Phi)$ for the energy band $m$ is well described by:
\begin{align}
	E_m = (E_L/4)\mathcal{M}_A&(m+1-(m+1)[\mathrm{mod}\,2]\\
	\nonumber & +2(-1)^m\Phi/\Phi_0;\,-E_S(\rho)/E_L)\,,
\end{align}
where $\mathcal{M}_A$ is the Mathieu characteristic function \cite{PhDThesis.Cottet}. The critical voltage for the SFT can be obtained as:
\begin{equation}
	\label{eqn:Vc}
	V_{C}(\Phi)=\min_{m}\left[\max_\rho\left.\frac{\pi}{e}\frac{\partial E_m(\rho,\Phi)}{\partial \rho}\right|_\Phi\right]\,.
\end{equation}
The critical voltage is the periodic function of an external magnetic flux $\Phi$ with a period of $\Phi_0$. The current-voltage characteristics (IVC) for the voltage biased SFT are similar to the characteristics of a single CQPS junction, with the same asymptotic resistance $R$. When the Stewart-McCumber parameter $\beta=(\pi L_\Sigma)/(eR^2)V_C\ll 1$ the IVC is given by:
\begin{equation}
	\label{eqn:IV}
	I=\sqrt{V^2-V_C^2(\Phi)}/R.
\end{equation}
The experiments reported here were performed in a $\mathrm{^{3}He}$ - $\mathrm{^{4}He}$ dilution refrigerator with a base temperature of about $\mathrm{20\,mK}$. The cryostat was equipped with a superconducting $\mathrm{5\,T}$ magnet. Signal leads were filtered by using a combination of lumped-element filters and a $\mathrm{ 1.5\,m }$ long Thermocoax cable between the $\mathrm{1\,K}$ stage and the RF-tight sample holder, mounted at the base temperature. To realize the pure voltage mode for our device, we placed a metallic ground plane under the formed structures \cite{PhysRevLett.105.026803.Pekola}. The ground plane consisted of a conductive $\mathrm{100\,nm}$ thick gold layer covered with a $\mathrm{100\,nm}$ thick layer of high-quality amorphous aluminum oxide ($\mathrm{Al O_x}$) insulating film. The SFT structures were patterned by conventional soft-mask electron-beam lithography on top of the $\mathrm{Al O_x}$. The final structures were formed by the successive e-gun deposition of highly disordered $\mathrm{ 40\,nm }$ amorphous indium oxide ($\mathrm{InO_x}$) film. Amorphous $\mathrm{InO_x}$ films exhibit a disorder-driven superconducting-insulator transition. The level of disorder can be controllably decreased by annealing at $\mathrm{40\,^\circ C}$ in vacuum \cite{PhysRevB.28.5075.Fiory}.

\begin{figure}
\centering
\includegraphics[width = \linewidth]{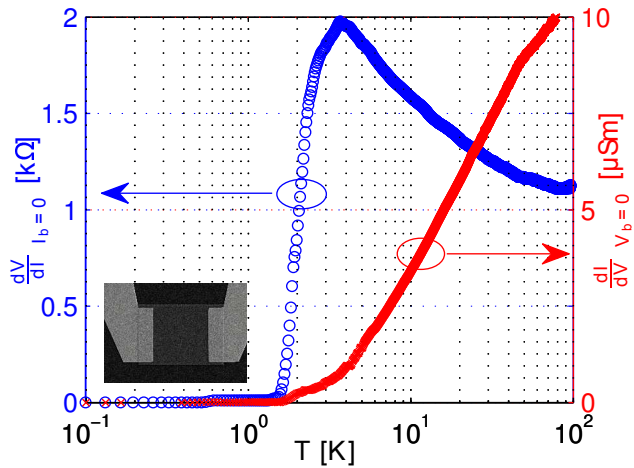}
\caption{\label{fig:TempDependence}
Data ($\circ$) corresponding to the left axis (blue color online)show the temperature dependence of the zero bias differential resistance ($\left.dV/dI\right|_{I_b=0}$) for a $30\times 30\,\mathrm{\mu m^2}$ test sample, whose micrograph of which is shown in the inset. The superconducting transition temperature $T_c \approx 2\,\mathrm{K}$ was defined as the temperature at which the resistance reduces to half the maximum value. The temperature dependence of the zero bias differential conductance ($\left.dI/dV\right|_{V_b=0}$) of the SFT is shown in red ($\times$). This data were used for a calibration curve to obtain the bias dependence of the quasiparticle temperature for the SFT.}
\end{figure}
Repetitive annealing reduced the disorder of the $\mathrm{InO_x}$ films until it reached a level where the superconductivity appeared in the test 2D structure (see inset of Fig.\,\ref{fig:TempDependence}) fabricated on chip close to the investigated SFT. The typical temperature dependence of the differential resistance ($R=dV/dI|_0$) for the test structure is shown by ($\circ$) blue curve in Fig.\,\ref{fig:TempDependence}. In the vicinity of $T=3.6\,\mathrm{K}$ the film resistance reaches the maximum value $R_\square=1.97\,\mathrm{k\Omega}$. The superconducting transition temperature $T_c \approx 2\,\mathrm{K}$ was defined as the temperature at which the resistance falls to half of the $R_\square$ value. The superconducting transition has a width of $\Delta T \approx 2\,\mathrm{K}$. The observed broadening is an indication of the large fluctuations in the superconducting order parameter of these films. Data referring to the left axis in Fig.\,\ref{fig:TempDependence} show the temperature dependence of the differential conductance ($\sigma=dI/dV|_0$) for the original SFT structure. In contrast to the superconducting transition in the test structures, the differential conductance of the SFT continuously decreases with temperature decrease and becomes almost indistinguishable from zero, within the accuracy of our experiments, at temperatures below $200\,\mathrm{mK}$. Such a temperature dependence is expected for CQPS junctions. Clearly, the observed temperature dependence of the sheet resistance of the nanowire, which is opposite to the dependence measured for the 2D film of the same thickness fabricated on chip under the same conditions, is an indicator superconductor -- insulator transition.

\begin{figure}[b]
\centering
\includegraphics[width = \linewidth]{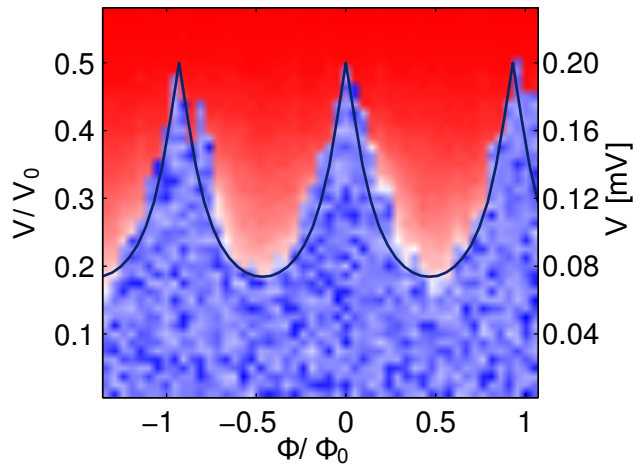}
\caption{\label{fig:FieldDep}
(color on-line) Experimentally measured stability diagram of the SFT on the plane of the bias voltage $V$, and the normalized external magnetic flux $\Phi/\Phi_0$. To emphasize the observed features we plot $\log|I|$; the red (blue) areas correspond to higher (lower) current with a range of  $I\in[0,\, 0.4]\,\mathrm{nA}$. The black solid line shows the flux dependence of the phase-slip critical voltage $V_C(\Phi)$ obtained from the model Eqs.(\ref{eqn:Hamiltonian1}-\ref{eqn:Vc}) for $E_S/E_L=0.4$. The left axis shows the normalized voltage $V/V_0$, where $V_0=(\pi/e)E_S$.}
\end{figure}
We investigated the IVC of the SFT at different temperatures and magnetic fields. Fig.\,\ref{fig:FieldDep} shows the experimentally measured stability diagram of the SFT in the plane of the normalized bias voltage $V$, and the normalized external magnetic flux $\Phi/\Phi_0$, with the color codes corresponding to the current through the device. To emphasize the observed features we plot $\log|I|$; the red (blue) areas correspond to higher (lower) current with the range of $I \in [0,\,0.4]\,\mathrm{nA}$. By assuming a symmetry in our device ($E_{S_1}=E_{S_2}$), the analytical expressions for the maximum and minimum values of the phase-slip blockade can be obtained in case of $E_L \gg E_S$. The maximum value occurs at the degeneracy point $\Phi/\Phi_0=(n+1/2)$, ($n\in\mathbb{Z}$), where $\max\{V_C\}=(\pi/e)(E_{S_1}/2)$. The minimum value $\min\{V_C\}\approx(\pi/e)(E_{S_1}^2/(2E_L))$ occurs, when the external flux is a multiple of the magnetic flux quantum $\Phi/\Phi_0=n$. Thus, from the experimental data we can extract the phase-slip energy $E_{S_1}\approx 0.11\,\mathrm{meV}$ of the CQPS junction in our device, and the magnetic energy $E_L\approx 0.27\,\mathrm{meV}$. The estimations of the kinetic inductance made in the BCS theory framework gives the same order of magnitude for the magnetic energy $E_L\approx 0.15\,\mathrm{meV}$. The black solid line on stability diagram shows the phase dependence of the critical voltage obtained from the described theoretical model Eqs.(\ref{eqn:Hamiltonian1}-\ref{eqn:Vc}). In these calculations the ratio between the phase slips and magnetic energies was set close to the experimentally observed value $E_{S}/E_L=0.4$. In spite of the good qualitative agreement between the experimental data and the described theoretical model, the phase-slip coupling energy obtained from the experiment can be smaller than the value predicted from the model \cite{Phys.Rep.464.1.Arutyunov}, since the observed critical voltage can be affected by the self-heating mechanism in our device. Similar behavior is observed in dual systems, namely Josephson weak links, where the supercurrent is always smaller than the maximum value predicted by the Ginzburg-Landau equations.

\begin{figure}
\centering
\includegraphics[width = \linewidth]{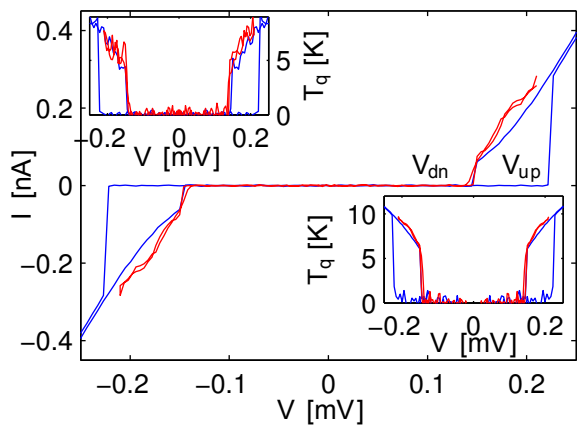}
\caption{\label{fig:IV} IVC of the SFT measured in the same magnetic field for different cryostat operating temperatures. Hysteretic IVC (blue curve online) was obtained at $30\,\mathrm{mK}$; non-hysteretic IVC (red curve online) was obtained at $100\,\mathrm{mK}$. The top inset shows the quasiparticle temperatures for different bias voltages obtained from the IVC and calibration data in Fig.\,\ref{fig:TempDependence}. The bottom inset shows the quasiparticle temperatures for the IVC, obtained from the model \cite{PhysRevLett.102.176802.Ovadia}.}
\end{figure}

An example of the IVC, which was obtained at $20\,\mathrm{mK}$, is shown in Fig.\,\ref{fig:IV}. The IVC has a strong hysteresis, which depends on the direction of the bias sweep. When the voltage increases from zero, the current and the differential conductance remain negligible. However, at a certain switching voltage $V_{up}$ the current and the differential conductance jump by several orders of magnitude and remain nonzero for the higher voltages. By decreasing the bias from values larger than $V_{up}$, the current and the differential conductance jump back to zero. But, this re-trapping takes place at voltages where $|V_{dn}|<|V_{up}|$. At temperatures above $100\,\mathrm{mK}$ the IVC becomes non-hysteretic. When the Stewart-McCumber parameter \cite{JApplPhys.39.3113.McCumber} $\beta\gg 1$, the hysteretic behavior of the IVC can be explained by the internal dynamics of the phase slips \cite{NaturePhysics.2.169.Mooij}. However, in our experiments, we have the opposite limit, $\beta\sim 10^{-3}$, in which such a hysteresis is not expected. The hysteresis in our system can be explained by a different process, which is often observed in systems with strongly decoupled phonons and quasiparticles. The Joule self-heating process plays an important role in such systems. It causes hysteresis in both conventional Josephson weak links \cite{PhysRevLett.101.067002.Courtois, PhysRevB.68.134515.Tinkham} and highly disordered $\mathrm{InO_x}$ films \cite{PhysRevLett.102.176802.Ovadia}.

A systematic study of this hysteresis is not a subject of this report, but a qualitative evaluation based on simple assumptions can be undertaken here. On the assumption that quasiparticles interact strongly with each other, the quasiparticle temperature $T_q$ can be introduced. It is necessary to note, that this temperature differs from the phonon temperature $T_{ph}$, which is given by the cryostat operating temperature. Another assumption is that the differential conductance is a function of quasiparticle temperature and is independent of the bias voltage. By using these assumptions, the bias dependence of quasiparticle temperature can be obtained from the measured IVC and the temperature dependence of the zero bias differential conductance. The top inset in Fig.\,\ref{fig:IV} shows the bias dependences of $T_q$, which correspond to the presented IVC. In the phase-slip blockade, $T_q$ remains almost constant; however, at a $V_{up}$ bias, quasiparticle temperature rises to a much higher value, and continues increasing because of the increased Joule power $\mathcal{P}=VI$. When the bias decreases, quasiparticle temperature follows the same curve, i.e., it remains elevated until it drops to the bath temperature at the re-trapping voltage $V_{dn}$. This demonstrates that the hysteresis in our SFT is governed by quasiparticle overheating. The quasiparticle temperature can be estimated in the model framework \cite{PhysRevLett.102.176802.Ovadia}. The Joule heat dissipates in a quasiparticle bath, but, due to the quasiparticle-phonon interaction some energy is removed to the photon bath. The steady state temperature is obtained as the solution to the heat balance equation:
\begin{equation}
\label{eqn:HeatBalance}
	\mathcal{P}=\Sigma \mathbb{V}(T_{q}^6-T_{ph}^6)\,,
\end{equation}
where $\mathbb{V}$ is the sample volume and $\Sigma$ is the quasiparticle-phonon coupling constant. The bottom inset in Fig.\,\ref{fig:IV} shows the $T_q$ bias dependence as a solution of Eq.(\ref{eqn:HeatBalance}). For the $100\,\mathrm{mK}$ phonon temperature, this equation has a single solution. However, for $30\,\mathrm{mK}$ phonon temperature, the equation becomes bistable, where the two values quasiparticle temperature of provide the stable solution. The authors of Ref.\,\cite{PhysRevLett.102.176802.Ovadia} suggested that this bistability is the origin of the hysteresis in the IVC.

In conclusion, we have presented the transport measurements and theoretical description of the new device: coherent quantum phase slip single flux transistor. This device consists of two superconducting nanowires embedded in a superconducting ring. We demonstrated that the transport properties of this device are dual to those of the Cooper-pair transistor. Current-voltage characteristics show the periodic modulation of phase-slip blockade by an external magnetic field, with a period equal to the flux quantum.  We emphasize that the hysteretic behavior of IVC at the temperatures below $100\,\mathrm{mK}$ is associated with the internal self-heating mechanism.

We are grateful to O.\,J.\,T. Peltonen, Yu.\,A. Pashkin for the useful discussions. This work was supported RFBR \#13-02-91177, \#13-02-00579, the Grant of President of Russian Federation for support of Leading Scientific Schools \#6170.2012.2, RAS presidium and Russian Federal Government programs.

\end{document}